\documentclass[final,5p,times,twocolumn,sort&compress]{elsarticle}
\usepackage{graphicx}
\usepackage{dcolumn}
\usepackage{bm}
\usepackage{amsmath, amssymb}

\usepackage{color}
\usepackage{colortbl}

\journal{Physics Letters A}
\begin{document}

\begin{frontmatter}
  \title{High order three part split symplectic integrators: \\
    Efficient techniques for the long time simulation of the \\
    disordered discrete nonlinear Schr\"{o}dinger equation}
\author[aristotle,capetown]{Ch.~Skokos} \ead{haris.skokos@uct.ac.za}
\address[aristotle]{Physics
  Department, Aristotle University of Thessaloniki, GR-54124,
  Thessaloniki, Greece} \address[capetown]{Department of
    Mathematics and Applied Mathematics, University of Cape Town,
    Rondebosch, 7701, South Africa}
\author[lohrmann]{E.~Gerlach} \address[lohrmann]{Lohrmann Observatory,
  Technical University Dresden, D-01062, Dresden, Germany}
\author[massey]{J.D.~Bodyfelt} \ead{J.Bodyfelt@massey.ac.nz}
\address[massey]{Centre for Theoretical Chemistry \& Physics, The New
  Zealand Institute for Advanced Study, \\ Massey University Albany,
  Private Bag 102904, North Shore City, Auckland 0745, New Zealand}
\author[kent]{G.~Papamikos} \address[kent]{School of Mathematics,
  Statistics and Actuarial Science, University of Kent, Canterbury,
  CT2 7NF, UK}
\author[imcce]{S.~Eggl} \address[imcce]{IMCCE, Observatoire de Paris,
  77 Avenue Denfert-Rochereau, F-75014, Paris, France}
\begin{abstract}
  While symplectic integration methods based on operator splitting are
  well established in many branches of science, high order methods for
  Hamiltonian systems that split in more than two parts have not been
  studied in great detail. Here, we present several high
  order symplectic integrators for Hamiltonian systems that can be
  split in exactly three integrable parts. We apply these techniques,
  as a practical case, for the integration of the disordered, discrete
  nonlinear Schr\"{o}dinger equation (DDNLS) and compare
    their efficiencies. Three part split algorithms provide effective
  means to numerically study the asymptotic behavior of wave packet
  spreading in the DDNLS - a hotly debated subject in current
  scientific literature.
\end{abstract}

\begin{keyword}
Symplectic integrators \sep three part split \sep disorder \sep nonlinear
Schr\"{o}dinger equation \sep multidimensional Hamiltonian systems
\end{keyword}
\end{frontmatter}

\section{Introduction}\label{sec:int}

Following the time evolution of a dynamical system is generally
accomplished by solving its corresponding equations of motion. If, for
instance, the system under consideration can be described by an
autonomous Hamiltonian function $H(\vec{q}, \vec{p})$, with $\vec{q}$,
$\vec{p}$ respectively being vectors of the generalized coordinates
and momenta, the equations of motion can be readily derived via
Hamilton's equations.  One then attempts to determine the solution
$\vec{x}(t)= (\vec{q}(t), \vec{p}(t))$, $t>0$, for any given initial
condition $\vec{x}(0)$.  Formally this solution can be described by
the action of the operator $e^{t L_H}$, with $L_H= \sum_i
H_{p_i}\partial_{q_i}-H_{q_i}\partial_{p_i} $, on the initial
condition, i.e.~$\vec{x}(t)=e^{t L_H}\vec{x}(0)$. The Hamiltonian is
said to be integrable if the action of this operator is known
explicitly and the solution of the Hamilton equations of motion can be
written in a closed, analytic form. Unfortunately, this task is rarely
possible, but in most cases the true solution can be approximated
numerically.  General purpose numerical integration methods for
ordinary differential equations are capable of providing such
approximations.

In this respect, the so-called symplectic integration techniques are
of particular interest, as they are explicitly designed for the
integration of Hamiltonian systems (see, for example, Chap.~VI of
\cite{Hairer_etal_02}, \cite{SI_Ham, SI_Forest} and references
therein).  Assume that $H(\vec{q}, \vec{p})$ can be written as
$H(\vec{q}, \vec{p})=A(\vec{q}, \vec{p})+B(\vec{q}, \vec{p})$, so that
the action of operators $e^{t L_A}$ and $e^{t L_B}$ is known, and the
solution of their Hamilton equations of motion can be written
analytically, while $e^{\tau L_H}$ does not permit a closed analytical
solution of its equations of motion.  Then, a symplectic scheme for
integrating the equations of motion from time $t$ to time $t+\tau$
consists of approximating the operator $e^{\tau L_H}=e^{\tau
  (L_A+L_B)}$ by a product of $j$ operators $e^{c_i \tau L_A}$ and
$e^{d_i \tau L_B}$, which represent exact integrations of Hamiltonians
$A(\vec{q}, \vec{p})$ and $B(\vec{q}, \vec{p})$ over times $c_i \tau$
and $d_i \tau$ respectively, i.e.~$e^{\tau L_H}=\prod_{i=1}^j e^{c_i
  \tau L_A} e^{d_i \tau L_B} + \mathcal{O}(\tau^{n+1})$.  The
constants $c_i$ and $d_i$ are appropriately chosen to increase the
order of the remainder of this approximation. In practice, using this
symplectic integrator (SI) we approximate the dynamics of the real
Hamiltonian $H=A+B$ by a new one, $K=A+B+\mathcal{O}(\tau^n)$,
introducing an error term of order $\tau^n$ in each integration step
-- the SI is then said to be of order $n$.

By their construction SIs preserve the symplectic nature of the
Hamiltonian system and keep bounded the error of the computed value of
$H$ (which is an integral of the system, commonly referred as the
`energy') irrespectively of the total integration time. Generally,
this is not the case with non-symplectic integration algorithms.
Furthermore, many SIs permit the use of relatively large integration time
steps $\tau$ for acceptable levels of energy accuracy, resulting in
lower CPU time requirements. Due to these benefits, SIs became a
standard technique in Hamiltonian dynamics with particular importance
in long time integrations of multidimensional systems. Several SIs of
different orders based on this operator splitting have been developed
over the years by various researchers \cite{Y90, SI_FR90,
  SI_CR91, SI_MA92, SI_Y93, SI_M95, SI_C97, BCFLMM13, FLBCMM13, LR01}.

\section{Three Part Split Symplectic Integrators}\label{sec:3}

In many cases the Hamiltonian can be written as a sum of the system's
kinetic energy $T(\vec{p})$, dependent only on the momenta $\vec{p}$,
and the potential $V(\vec{q})$, dependent only on the positions
$\vec{q}$. Then the obvious choice for the application of a SI is to
consider $A \equiv T(\vec{p})$ and $B \equiv V(\vec{q})$. Yet in many
physical problems, the corresponding Hamiltonian cannot be split in
two integrable parts -- is it possible to exploit the advantages of
SIs for such systems as well? The answer to this question is positive
as, theoretically, symplectic integration schemes can be constructed
for Hamiltonian systems that split in an arbitrary number of
integrable parts
\cite{CN96}\cite[][Sect.~II.5]{Hairer_etal_02}. Of course
  the construction of high order SIs is not an easy task as the number
  of involved operators increase extremely fast. This problem becomes
  even more complicated when the Hamiltonian is split in three,
  instead of two, integrable parts.

In this paper we systematically present and test the
 performance of efficient high order SIs for Hamiltonians that can
be split in three integrable parts. Particular cases of second order
three part split SIs, connected with astronomical problems, have been
reported in literature \cite{SI_C99, SI_GBB08, SI_QPRB10, SI_C10}. In
these works, the considered Hamiltonians were expressed as
$H=A(\vec{q}, \vec{p})+B(\vec{q}, \vec{p})+C(\vec{q},\vec{p})$, the
action of operators $e^{\tau L_A}$, $e^{\tau L_B}$ and $e^{\tau L_C}$
was analytically obtained, the second order SI of 5 steps
\begin{equation}\label{eq:ABC_2}
ABC^2(\tau)=e^{\frac{\tau}{2} L_A} e^{\frac{\tau}{2} L_B}  e^{\tau L_C}
e^{\frac{\tau}{2} L_B} e^{\frac{\tau}{2} L_A}
\end{equation}
was constructed, and its performance was studied. This integrator
represents the simplest form of a symmetric SI that can be constructed
for a Hamiltonian which splits in three distinct parts,
  as was also explained in \cite{K96}.

Some attempts to create three part split SIs of order higher than two
can be found in the literature. In \cite{K96} an
  integrator of order four was obtained, while in \cite{O_07} second
and fourth order integration schemes for a particular complicated
molecular model were presented. Recently, in
  \cite{BCFLMM13,FLBCMM13} three part split SIs especially oriented
  for near integrable Hamiltonians of the form $H=A+\epsilon(B+C)$,
  with $\epsilon \ll 1$, were constructed and applied to a specific
  astronomical problem. In principle these integrators can be applied
  to any Hamiltonian that split in three integrable parts, and we will
  use some of them later on in Sect.~\ref{sec:alt}.

A general way to obtain higher order SIs is the construction of
symmetric compositions of a basic symmetric second order
integrator. The number of times that this basic integrator is used in
a particular scheme determines the number of `\emph{stages}' of the
constructed integrator. This approach led to the creation of efficient
schemes of order six, eight and ten \cite{KL97,SS05} (see also
\cite{BCM08} for a detailed review of these methods), but to the best
of our knowledge, it has never been systematically applied to
Hamiltonians that split in three integrable parts.

\subsection{Integrators of Order Four}\label{sec:ord4}

We start the presentation of three part split methods by implementing
an algorithm based on the composition technique proposed by Yoshida
\cite{Y90}. Starting from a SI $S^{2n}(\tau)$ of order $2n$, we can
construct a SI $S^{2n+2}(\tau)$ of order $2n+2$, as
\begin{equation}\label{eq:CompYosh}
    S^{2n+2}(\tau)=S^{2n}(z_1\tau)S^{2n}(z_0\tau)S^{2n}(z_1\tau),
\end{equation}
with \mbox{$z_0=-2^{1/(2n+1)} / [2-2^{1/(2n+1)}]$} and \mbox{$z_1=1
  /[2-2^{1/(2n+1)}]$}.  Applying this procedure to the second order SI
(\ref{eq:ABC_2}) we obtain the fourth order SI of 3
  stages and 13 steps
\begin{equation}\label{eq:ABC_4}
ABC^4_{[\text{Y}]}(\tau)=ABC^2(x_1\tau)ABC^2(x_0\tau)ABC^2(x_1\tau),
\end{equation}
with

\begin{equation}\label{eq:ABC_4_coef}
x_0=\frac{-\sqrt[3]{2}}{2-\sqrt[3]{2}}, \,\,\,\, x_1=\frac{1}{2-\sqrt[3]{2}},
\end{equation}
and the subscript [Y] referring to the author of \cite{Y90}. We note
that the $ABC^4_{[\text{Y}]}$ was explicitly constructed in
\cite{K96}.

We also consider another composition scheme which was introduced in
\cite{S90} and studied in \cite{KL97} (where it was named `s5odr4')
and \cite{SS05}. This scheme has 5 stages and starting from a second
order SI, which in our case will be the $ABC^2$ integrator
(\ref{eq:ABC_2}), leads to the fourth order integrator
\begin{equation}\label{eq:ABC_4S}
    \begin{array}{l}
       ABC^4_{[\text{S}]}(\tau)=ABC^2(p_2\tau)ABC^2(p_2\tau)
    ABC^2((1-4p_2)\tau) \times \\
       \,\,\,\,  \times ABC^2(p_2\tau)ABC^2(p_2\tau),
    \end{array}
\end{equation}
with
\begin{equation}\label{eq:ABC_4S_coef}
    p_2=\frac{1}{4-\sqrt[3]{4}}, \,\,\,\,
    1-4p_2=-\frac{\sqrt[3]{4}}{4-\sqrt[3]{4}}.
\end{equation}
which has 21 steps. As in the previous case the subscript [S] refers
to the author of \cite{S90}.

\subsection{Integrators of Order Six}\label{sec:ord6}

Equation (\ref{eq:CompYosh}) can be used repeatedly to
  get higher order three part split SIs. Although such a procedure
for  obtaining arbitrary SIs of even order with exact
coefficients is straightforward, it is not optimal with respect to the
number of required steps. As was already pointed out in \cite{Y90},
alternative methods can be applied to obtain more economical
integrators of high order, although the new coefficients can no longer
be given in analytical form. Several sixth order SIs of this kind were
presented in \cite{Y90}. Here, we consider one corresponding to
`solution A' in \cite{Y90}
\begin{equation}\label{eq:ABC_6}
    \begin{array}{l}
      ABC^6_{[\text{Y}]}(\tau)=ABC^2(w_3\tau)ABC^2(w_2\tau)ABC^2(w_1\tau)
      \times \\ \,\,\,\, \times
      ABC^2(w_0\tau)ABC^2(w_1\tau)ABC^2(w_2\tau)ABC^2(w_3\tau)
    \end{array}
\end{equation}
having 7 stages and 29 steps. The exact values of $w_i$,
$i=0,1,2,3$ can be found in \cite[Chap.~V, Eq.~(3.11)]{Hairer_etal_02}
and \cite{Y90}.  We include this particular integrator in
  our study because according to \cite{M95} it shows the best behavior
  among the ones presented in \cite{Y90}. We also note that this
  integrator corresponds to the `s7odr6' method studied in
  \cite{KL97}.

In addition we include in our study other SIs of order six obtained by
composition techniques which involve more stages than the
ABC$^6_{[\text{Y}]}$ integrator. In particular we consider the
`s9odr6b' integrator of \cite{KL97} which has 9 stages, i.e.~9
implementations of a second order SI. Using the $ABC^2$ method
(\ref{eq:ABC_2}) as such an integrator we end up with the scheme
\begin{equation}\label{eq:ABC_6KL}
    \begin{array}{l}
      ABC^6_{[\text{KL}]}(\tau)=
      ABC^2(\delta_1\tau)ABC^2(\delta_2\tau)ABC^2(\delta_3\tau)
      ABC^2(\delta_4\tau) \times \\ \,\,\,\, \times
      ABC^2(\delta_5\tau) ABC^2(\delta_4\tau) ABC^2(\delta_3\tau)
      ABC^2(\delta_2\tau)ABC^2(\delta_1\tau)
    \end{array}
\end{equation}
of 37 steps. We note that the subscript [KL] refers to the initials of
the authors of \cite{KL97}, and the exact values of constants
$\delta_i$, $1\leq i \leq 5$ are given in the appendix of \cite{KL97}.

We also consider a sixth order SI based on a composition
method with 11 stages, which was introduced in \cite{SS05}. This
approach leads to the SI
\begin{equation}\label{eq:ABC_6SS}
    \begin{array}{l}
      ABC^6_{[\text{SS}]}(\tau)=
      ABC^2(\gamma_1\tau)ABC^2(\gamma_2\tau)\cdots ABC^2(\gamma_5\tau)
      \times \\ \,\,\,\, \times ABC^2(\gamma_6\tau)ABC^2(\gamma_5\tau)
      \cdots ABC^2(\gamma_2\tau)ABC^2(\gamma_1\tau)
    \end{array}
\end{equation}
which has 45 individual steps. Again the subscript [SS] refers to the
authors of \cite{SS05}, while the exact values of $\gamma_i$, $1\leq i
\leq 6$ are given in Eq.~(11) of \cite{SS05}.  

\subsection{Integrators of Order Eight}\label{sec:ord8}

In \cite{Y90} five different composition methods of 15 stages that
lead to eighth order SIs are given. Among them the one named `solution
D' exhibits the best behavior according to \cite{M95,SS05}. For this
reason we include this composition method in our study. The resulting
SI (using the constants $w_i$, $0 \leq i \leq 7$ appearing in Table 2
of \cite{Y90}) is
\begin{equation}\label{eq:ABC_8}
    \begin{array}{l}
      ABC^8_{[\text{Y}]}(\tau)= ABC^2(w_7\tau)ABC^2(w_6\tau)\cdots
      ABC^2(w_1\tau) \times \\ \,\,\,\, \times ABC^2(w_0\tau)
      ABC^2(w_1\tau) \cdots ABC^2(w_6\tau) ABC^2(w_7\tau)
    \end{array}
\end{equation}
having 61 individual steps.

We also consider two more SIs of order eight obtained by composition
techniques which involve more stages than the ABC$^8_{[\text{Y}]}$
integrator. The first is based on the `s17odr8b' integrator of
\cite{KL97} which has 17 stages. Its form is
\begin{equation}\label{eq:ABC_8KL}
    \begin{array}{l}
      ABC^8_{[\text{KL}]}(\tau)=
      ABC^2(\delta_1\tau)ABC^2(\delta_2\tau)\cdots ABC^2(\delta_8\tau)
      \times \\ \,\,\,\, \times ABC^2(\delta_9\tau)
      ABC^2(\delta_8\tau) \cdots
      ABC^2(\delta_2\tau)ABC^2(\delta_1\tau).
    \end{array}
\end{equation}
This integrator has 69 steps and its coefficients can be found in the
appendix of \cite{KL97}. The second integrator is
\begin{equation}\label{eq:ABC_8SS}
    \begin{array}{l}
      ABC^8_{[\text{SS}]}(\tau)=
      ABC^2(\gamma_1\tau)ABC^2(\gamma_2\tau)\cdots ABC^2(\gamma_9\tau)
      \times \\ \,\,\,\, \times ABC^2(\gamma_{10}\tau)
      ABC^2(\gamma_9\tau) \cdots
      ABC^2(\gamma_2\tau)ABC^2(\gamma_1\tau)
    \end{array}
\end{equation}
and is based on a composition method with 19 stages presented in
Eq.~(13) of \cite{SS05}.

\subsection{An Integrator of Order Ten}\label{sec:ord10}

Finally, as an extreme case, we include in our study a SI of order
ten. In particular we consider the tenth order composition method of
31 stages presented in Eq.~(15) of \cite{SS05}, which leads to the SI
\begin{equation}\label{eq:ABC_10SS}
    \begin{array}{l}
      ABC^{10}_{[\text{SS}]}(\tau)=
      ABC^2(\gamma_1\tau)ABC^2(\gamma_2\tau)\cdots
      ABC^2(\gamma_{15}\tau) \times \\ \,\,\,\, \times
      ABC^2(\gamma_16\tau) ABC^2(\gamma_{15}\tau) \cdots
      ABC^2(\gamma_2\tau)ABC^2(\gamma_1\tau)
    \end{array}
\end{equation}
with 125 steps. We choose to not include additional integrators of
order ten based on compositions techniques with more stages due to the
substantial increase of their complexity.

In Table \ref{tab:ABC} we present all the three part split SIs used in
our study providing information about their order, the number of their
stages and steps, as well as references for obtaining the values of
their coefficients.

\begin{table*}
\centering
\begin{tabular}{|c|c|c|c|c|}
  \hline
 \rowcolor[gray]{0.9} \textbf{SI} & \textbf{Order} & \textbf{Stages} & \textbf{Steps} & \textbf{Coefficients} \\
\hline
  $ABC^2$ & 2 & 1 & 5 & (\ref{eq:ABC_2}) \\
\hline
  $ABC^4_{[\text{Y}]}$ & 4 & 3 & 13 & (\ref{eq:ABC_4_coef}) \\
\hline
  $ABC^4_{[\text{S}]}$ & 4 & 5 & 21 & (\ref{eq:ABC_4S_coef}) \\
\hline
  $ABC^6_{[\text{Y}]}$ & 6 & 7 & 29 &  `Solution A' in Table 1 of
\cite{Y90}\\
\hline
  $ABC^6_{[\text{KL}]}$ & 6 & 9 & 37 &  Table 
`s9odr6b' in the appendix of \cite{KL97}\\
\hline
  $ABC^6_{[\text{SS}]}$ & 6 & 11 & 45 &  Equation (11) of \cite{SS05}\\
\hline
  $ABC^8_{[\text{Y}]}$ & 8 & 15 & 61 &  `Solution D' in Table 2 of
\cite{Y90}\\
\hline
  $ABC^8_{[\text{KL}]}$ & 8 & 17 & 69 &  Table `s17odr8b' 
in the appendix of \cite{KL97}\\
\hline
  $ABC^8_{[\text{SS}]}$ & 8 & 19 & 77 &  Equation (13) of \cite{SS05}\\
\hline
  $ABC^{10}_{[\text{SS}]}$ & 10 & 31 & 125 &  Equation (15) of \cite{SS05}\\
  \hline
\end{tabular}
\caption{Information for the three part split SIs of
    Sect.~\ref{sec:3}. For each integrator we provide its name, its
    order, the number of its stages (i.e.~the appearances of the
    second order SI $ABC^2$ (\ref{eq:ABC_2})) and the total number of
    individual steps. In the last column (named `Coefficients') we
    indicate where the explicit values of the coefficients appearing
    in each step can be found. For example (\ref{eq:ABC_4_coef})
    refers to Eq.~(\ref{eq:ABC_4_coef}) of this paper.}
\label{tab:ABC}
\end{table*}

\section{Integration of the Disordered Discrete Nonlinear Schr\"{o}dinger
Equation}\label{sec:DDNLS}

In order to investigate the efficiency of the different SI schemes we
choose a multidimensional Hamiltonian system describing a
one--dimensional chain of coupled, nonlinear oscillators. In
particular we consider the Hamiltonian of the disordered discrete
nonlinear Schr\"odinger equation (DDNLS)
\begin{equation}
\mathcal{H}_{D}= \sum_{l} \epsilon_{l}
|\psi_{l}|^2+\frac{\beta}{2} |\psi_{l}|^{4}
- (\psi_{l+1}\psi_l^*  +\psi_{l+1}^* \psi_l),
\label{RDDNLS}
\end{equation}
with complex variables $\psi_{l}$, lattice site indices $l$ and
nonlinearity strength $\beta \geq 0$.  The random on--site energies
$\epsilon_{l}$ are chosen uniformly from the interval
$\left[-\frac{W}{2},\frac{W}{2}\right]$, with $W$ denoting the
disorder strength. This model has two integrals of motion, as it
conserves both the energy (\ref{RDDNLS}) and the norm $S =
\sum_{l}|\psi_l|^2$, and has been extensively investigated in order to
determine the characteristics of energy spreading in disordered
systems \cite{KKFA08, BLGKSF11,FKS09,SKKF09,LBKSF10,BLSKF11}. These
studies showed that the second moment, $m_2$, of the norm distribution
grows subdiffusively in time $t$, as $t^a$, and the asymptotic value
$a=1/3$ of the exponent was theoretically predicted and numerically
verified.  Currently open questions on the dynamics of disordered
systems concern the possible halt of wave packet's spreading for
$t\rightarrow \infty$ \cite{JKA10, A11}, as well as the
characteristics of its chaotic behavior.  Thus, providing the means to
perform accurate long time simulations for the DDNLS model within
reasonable amounts of computational time is essential.

Applying the canonical transformation
\mbox{$\psi_l=(q_l+ip_l)/\sqrt{2}$},
\mbox{$\psi_l^*=(q_l-ip_l)/\sqrt{2}$}, one can split (\ref{RDDNLS})
into a sum of there integrable parts $A$, $B$ and $C$ as follows
\begin{equation}
H_D=\sum_l\underbrace{\tfrac{\epsilon_l}{2}(q_l^2+p_l^2)+\tfrac{\beta}{8}
(q_l^2+p_l^2)^2}_A \underbrace{- p_{l+1} p_l}_B \underbrace{-q_{l+1}q_l}_C,
\label{HDDNLS}
\end{equation}
where $q_l$ and $p_l$ are respectively generalized coordinates and
momenta. For these three parts the propagation of initial conditions
$(q_l, p_l)$ at time $t$, to their final values $(q'_l, p'_l)$ at time
$t+\tau$ is given by the operators
\begin{equation}
e^{\tau L_A}: \left\{ \begin{array}{lll} q'_l & = & q_l \cos(\alpha_l
  \tau)+ p_l \sin(\alpha_l \tau)\\ p'_l & =& p_l \cos(\alpha_l \tau)-
  q_l \sin(\alpha_l \tau) \\
\end{array}\right. ,
\label{eq:LA}
\end{equation}
\begin{equation}
e^{\tau L_B}: \left\{ \begin{array}{lll} p'_l & =& p_l \\ q'_l & = &
q_l-(p_{l-1}+p_{l+1}) \tau \\
\end{array}\right. ,
\label{eq:LB}
\end{equation}
\begin{equation}
e^{\tau L_C}: \left\{ \begin{array}{lll} q'_l & =& q_l \\ p'_l & = &
p_l+(q_{l-1}+q_{l+1}) \tau \\
\end{array}\right. ,
\label{eq:LC}
\end{equation}
with $\alpha_l=\epsilon_l+\beta(q_l^2+p_l^2)/2$. Thus, the DDNLS model
represents an ideal test case for our aforementioned three part split
SIs.

\subsection{Alternative Integration Approaches \label{sec:alt}}

In order to evaluate the efficiency of the integration
  schemes presented in Sect.~\ref{sec:3}, we compare their performance
to that of other numerical techniques. In
\cite{FKS09,SKKF09,LBKSF10,BLSKF11} numerical integration schemes
based on traditional two part split SIs were applied for the
integration of Hamiltonian (\ref{HDDNLS}).  These approaches were
based on the split of (\ref{HDDNLS}) in two parts as $\mathcal{A}=A$
and $\mathcal{B}=B+C$, and the application of second order SIs of the
so--called SABA--family \cite{LR01}; note that the SABA$_1$ integrator
is more popularly known as the St\"{o}rmer-Verlet leapfrog integrator.

In our study we implement the
second order SI SABA$_2$ using the split
$H_D=\mathcal{A}+\mathcal{B}$. The integration of the $\mathcal{A}$
part is performed according to (\ref{eq:LA}), while different
approaches for approximating the action of $e^{\tau
  L_{\mathcal{B}}}=e^{\tau L_{B+C}}$ are followed.  In
\cite{FKS09,SKKF09} a numerical scheme based on Fourier transforms was
implemented (see appendix of \cite{SKKF09} for more details) leading
to a second order integrator with 5 steps, which we name SIFT$^2$ in
the following. Another approach is to split the $\mathcal{B}$ part in
two integrable parts as $\mathcal{B}=B+C$ and use the SABA$_2$ SI to
approximate its solution. This means that we perform two successive
two part splits in order to integrate $H_D$. This approach leads to a
second order SI with 13 steps which we name SS$^2$ (this scheme
corresponds to the PQ method used in \cite{BLSKF11}).

Extending the approach to split $H_D$ (\ref{HDDNLS}) in
  two parts where the $\mathcal{A}=A$ is integrable and the
  $\mathcal{B}=B+C$ part is approximately integrated either by another
  two part split SI or by an appropriate Fourier transform scheme, we
  construct fourth order integrators, which, to the best of our
  knowledge, have never being used before for the integration of the
  DDNLS system. In particular, by applying the composition procedure
  (\ref{eq:CompYosh}) to the SS$^2$ integrator we construct a fourth
  order integrator with 37 simple steps that we call SS$^4$. Following
  a similar approach for the SIFT$^2$ integrator we obtain a fourth
  order integrator with 13 steps, which we name SIFT$^4$.

In addition, we use some recently introduced SIs
\cite{BCFLMM13,FLBCMM13} which were particularly constructed for
nearly integrable Hamiltonians, i.e. Hamiltonians of the form
$H=\mathcal{A}+\epsilon \mathcal{B}$, where the $\mathcal{A}=A$ part
is integrable and $\epsilon \ll 1$.  In particular, we consider the
fourth order integrators ABA864, ABA1064, ABAH864, ABAH1064, where the
$\mathcal{A}$ part is integrated explicitly and the $\mathcal{B}$ part
either by the Fourier transforms (for the ABA864, ABA1064 integrators)
or by the SABA$_2$ SI (for the ABAH864, ABAH1064 integrators). We note
that the ABAH864 and ABAH1064 schemes were constructed from the ABA864
and ABA1064 integrators respectively, by assuming that the
$\mathcal{B}$ part is a second order symmetric integrator
\cite{BCFLMM13} (which in our study is the SABA$_2$ scheme). This
assumption leads to an additional condition of the integrator's
coefficients, which in turn results to the addition of some more steps
in the integrator. As the solution of the $\mathcal{B}$ part by
Fourier transforms is a rather time consuming procedure, we decided to
use this approach for solving the $\mathcal{B}$ part when applying the
ABA864 and ABA1064 methods which have less individual steps.

In particular, based on 
 the ABA864 and ABA1064 integrators of
\cite{BCFLMM13} we consider the fourth order schemes
\begin{equation}\label{eq:SIFT_4_864}
    \begin{array}{l}
      SIFT^4_{864}(\tau)= e^{\alpha_1 L_{A}} e^{b_1 L_{\mathcal{B}}}
      e^{\alpha_2 \tau L_{A}} e^{b_2 \tau L_{\mathcal{B}}} \cdots
      e^{\alpha_4 \tau L_{A}} e^{b_4 \tau L_{\mathcal{B}}} \times
      \\ \,\,\,\, \times e^{\alpha_4 \tau L_{A}} e^{b_3
        L_{\mathcal{B}}} e^{\alpha_3 \tau L_{A}} e^{b_2 \tau
        L_{\mathcal{B}}} e^{\alpha_2 \tau L_{A}} e^{b_1 \tau
        L_{\mathcal{B}}} e^{\alpha_1 \tau L_{A}}
    \end{array}
\end{equation}
and
\begin{equation}\label{eq:SIFT_4_1064}
    \begin{array}{l}
      SIFT^4_{1064}(\tau)= e^{\alpha_1 \tau L_{A}} e^{b_1 \tau
        L_{\mathcal{B}}} e^{\alpha_2 \tau L_{A}} e^{b_2 \tau
        L_{\mathcal{B}}} \cdots e^{\alpha_4 \tau L_{A}} e^{b_4 \tau
        L_{\mathcal{B}}} \times \\ \,\,\,\, \times e^{\alpha_5 \tau
        L_{A}} e^{b_4 \tau L_{\mathcal{B}}} e^{\alpha_4 \tau L_{A}}
      \cdots e^{b_2 \tau L_{\mathcal{B}}} e^{\alpha_2 \tau L_{A}}
      e^{b_1 \tau L_{\mathcal{B}}} e^{\alpha_1 \tau L_{A}},
    \end{array}
\end{equation}
with 43 and 49 steps respectively, where the $\mathcal{B}=B+C$ part is
integrated according to the Fourier transform procedure presented in
\cite{SKKF09}. The values of the coefficients appearing in
(\ref{eq:SIFT_4_864}) and (\ref{eq:SIFT_4_1064}) are given in Table 3
of \cite{BCFLMM13}.

Similarly, based on
 the ABAH864 and ABAH1064 integrators of
\cite{BCFLMM13} we consider
 the fourth order integrators
\begin{equation}\label{eq:SS_4_864}
    \begin{array}{l}
      SS^4_{864}(\tau)= e^{\alpha_1 \tau L_{A}} e^{b_1 \tau
        L_{\mathcal{B}}} e^{\alpha_2 \tau L_{A}} e^{b_2 \tau
        L_{\mathcal{B}}} \cdots e^{\alpha_4 \tau L_{A}} e^{b_4 \tau
        L_{\mathcal{B}}} \times \\ \,\,\,\, \times e^{\alpha_5 \tau
        L_{A}} e^{b_4 \tau L_{\mathcal{B}}} e^{\alpha_4 \tau L_{A}}
      \cdots e^{b_2 \tau L_{\mathcal{B}}} e^{\alpha_2 \tau L_{A}}
      e^{b_1 \tau L_{\mathcal{B}}} e^{\alpha_1 \tau L_{A}}
    \end{array}
\end{equation}
and
\begin{equation}\label{eq:SS_4_1064}
    \begin{array}{l}
      SS^4_{1064}(\tau)= e^{\alpha_1 \tau L_{A}} e^{b_1 \tau
        L_{\mathcal{B}}} e^{\alpha_2 \tau L_{A}} e^{b_2 \tau
        L_{\mathcal{B}}} \cdots e^{\alpha_5 \tau L_{A}} e^{b_5 \tau
        L_{\mathcal{B}}} \times \\ \,\,\,\, \times e^{\alpha_5 \tau
        L_{A}} e^{b_4 \tau L_{\mathcal{B}}} e^{\alpha_4 \tau
        L_{A}}\cdots e^{b_2 \tau L_{\mathcal{B}}} e^{\alpha_2 \tau
        L_{A}} e^{b_1 \tau L_{\mathcal{B}}} e^{\alpha_1 \tau L_{A}},
    \end{array}
\end{equation}
with 49 and 55 steps respectively, where the $\mathcal{B}=B+C$ part is
integrated by the SABA$_2$ SI. The values of the coefficients
appearing in (\ref{eq:SS_4_864}) and (\ref{eq:SS_4_1064}) are given in
Table 4 of \cite{BCFLMM13}.

Of course one can also use any general purpose non--symplectic
integrator for the integration of (\ref{HDDNLS}). One disadvantage of
such techniques is that different epochs of the system's evolution are
computed with different accuracy since these integrators do not keep
the energy error bounded, but increase it as time increases. In
particular for the DDNLS model considered here the later stages of its
evolution, which are of most importance since we are mainly interested
in the asymptotic behavior of the system, are computed less
accurately. As a representative of non--symplectic integrators we
consider here the variable step Runge--Kutta method called DOP853
\cite{DOP}, whose performance is controlled by the so--called
one--step accuracy $\delta$.

\section{Numerical Results}\label{sec:results}

In order to compare the performance of the various integration schemes
we consider a particular disorder realization of the DDNLS model
(\ref{HDDNLS}) with $N=1024$ lattice sites. We fix the total norm of
the system to $S=1$, and following \cite{LBKSF10} we initially excite
homogeneously 21 central sites by attributing to each one of them the
same constant norm, but with a random phase, while for all other sites
we set $q_l(0)=p_l(0)=0$. Due to the nonlinear nature of the model the
norm distribution spreads, keeping of course the total norm $S=\sum_l
(q_l^2+p_l^2)/2$ constant ($S=1$). The performance of the integration
schemes is evaluated by their ability to (a) reproduce correctly the
dynamics, which is reflected in the subdiffusive increase of $m_2(t)$,
(b) keep the values of the two integrals $H_D$, $S$ constant, as
monitored by the evolution of the absolute relative errors of the
energy $E_r(t)=|[H_D(t)-H_D(0)]/H_D(0)|$, and norm
$S_r(t)=|[S(t)-S(0)]/S(0)|$, and (c) reduce the required CPU time
$T_c(t)$ for the performed computations.

\begin{figure*}[t]
\includegraphics[width=\textwidth, keepaspectratio]{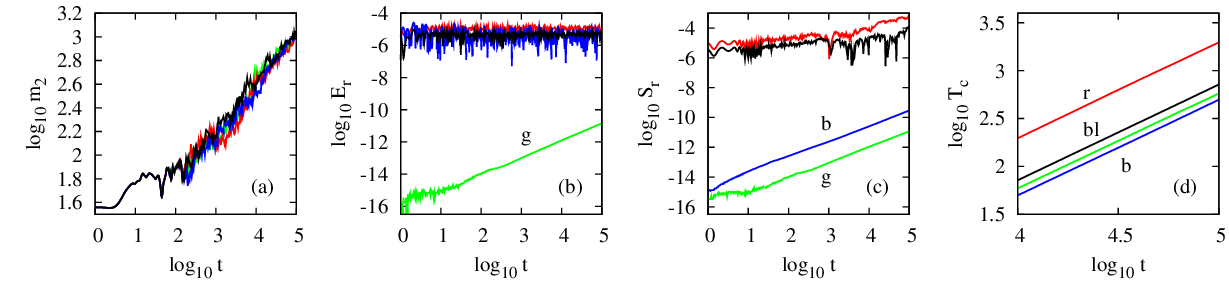}
\caption{(Color online) Results for the integration of
    $H_D$ (\ref{HDDNLS}) by the second order SIs ABC$^2$ for
    $\tau=0.005$, SS$^2$ for $\tau=0.02$, SIFT$^2$ for $\tau=0.05$
    [(r) red; (bl) black; (b) blue], and the non--symplectic
    integrator DOP853 for $\delta=10^{-16}$ [(g) green]: time
    evolution of the logarithm of (a) the second moment $m_2(t)$, (b)
    the absolute relative energy error $E_r(t)$, (c) the absolute
    relative norm error $S_r(t)$, and (d) the required CPU time
    $T_c(t)$ in seconds.}
\label{f1}
\end{figure*}

Results obtained by the second order SIs ABC$^2$, SS$^2$ and SIFT$^2$
and the non--symplectic integrator DOP853 are presented in
Fig.~\ref{f1}. These integration methods correctly describe the
system's dynamical evolution since for all of them the wave packet's
$m_2$ shows practically the same behavior (Fig.~\ref{f1}a).  The time
steps $\tau$ of the three SIs were chosen so that all of them keep the
relative energy error practically constant at $E_r\approx10^{-5}$
(Fig.~\ref{f1}b). Since we are interested in the accurate long time
integration of the DDNLS model we use $\delta=10^{-16}$ for the
implementation of the DOP853 integrator. For $t\approx 10^8$ (which
can be considered as a typical final integration time for long time
simulations), this choice results practically in the same energy error
obtained by all other tested integrators.  From Fig.~\ref{f1}c we see
that the relative norm error $S_r$ increases for all used methods,
exhibiting larger values yet lower increase rates, for the ABC$^2$ and
SS$^2$ SIs. Nevertheless, our results indicate that all methods can
keep $S_r$ to acceptable levels (e.g.~$S_r \lesssim 10^{-2}$), even
for long time integrations. It is worth noting that the
  Fourier transforms used by the SIFT$^2$ scheme for the integration
  of the $\mathcal{B}$ part of (\ref{HDDNLS}) preserve the norm $S$
  (see appendix of \cite{SKKF09} for more details).  For this reason
  the corresponding relative error $S_r$ attains smaller values than
  for the ABC$^2$ and SS$^2$ integrators (Fig.~\ref{f1}c).  From
Fig.~\ref{f1}d we see that the SIFT$^2$ integration scheme is the most
efficient one with respect to the CPU time needed for obtaining the
results of Fig.~\ref{f1}.

\begin{figure*}[t]
\includegraphics[width=\textwidth, keepaspectratio]{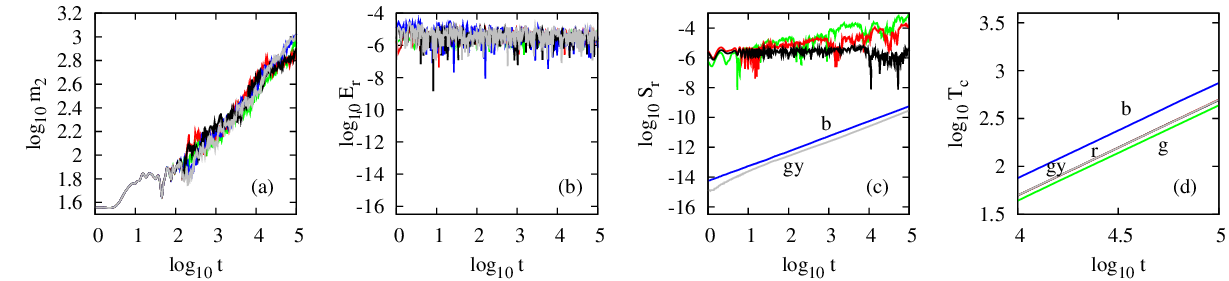}
\caption{(Color online) Results for the integration of
    $H_D$ (\ref{HDDNLS}) by the second order SI SIFT$^2$ for
    $\tau=0.05$ [(gy) grey], and the fourth order SIs
    ABC$^4_{[\text{Y}]}$ for $\tau=0.05$, ABC$^4_{[\text{S}]}$ for
    $\tau=0.1$, SIFT$^4$ for $\tau=0.125$ and SS$^4$ for $\tau=0.1$
    [(g) green; (r) red; (b) blue; (bl) black].  The panels are as in
    Fig.~\ref{f1}. Note that in panel (d) the red, blue and black
    curves practically overlap.}
\label{f2}
\end{figure*}

For this reason we use the SIFT$^2$ SI as a reference method, and
compare in Fig.~\ref{f2} its results with the ones obtained by
the fourth order SIs: ABC$^4_{[\text{Y}]}$,
  ABC$^4_{[\text{S}]}$, SIFT$^4$ and SS$^4$. These SIs reproduce
correctly the evolution of $m_2$ (Fig.~\ref{f2}a) and keep
$E_r\approx10^{-5}$ (Fig.~\ref{f2}b). $S_r$ for the
  SIFT$^4$ method shows a similar behavior to SIFT$^2$, while for all
  other integrators it attains larger, slowly increasing values, which
  nevertheless remain acceptably small (Fig.~\ref{f2}c). The SIFT$^4$
  method requires more CPU time than SIFT$^2$ (Fig.~\ref{f2}c),
  despite the fact it utilizes a larger time step, because it
  implements the CPU time consuming Fourier transforms more
  times. Consequently, the development of higher order schemes based
  on Fourier transforms for the integration of the $\mathcal{B}$ part
  of Hamiltonian (\ref{HDDNLS}) does not lead to very efficient
  schemes, with respect to CPU time requirements.  From the remaining
  integrators of Fig.~\ref{f2} the ABC$^4_{[\text{Y}]}$ requires the
  least CPU time (Fig.~\ref{f2}d).

\begin{figure*}[t]
\includegraphics[width=\textwidth, keepaspectratio]{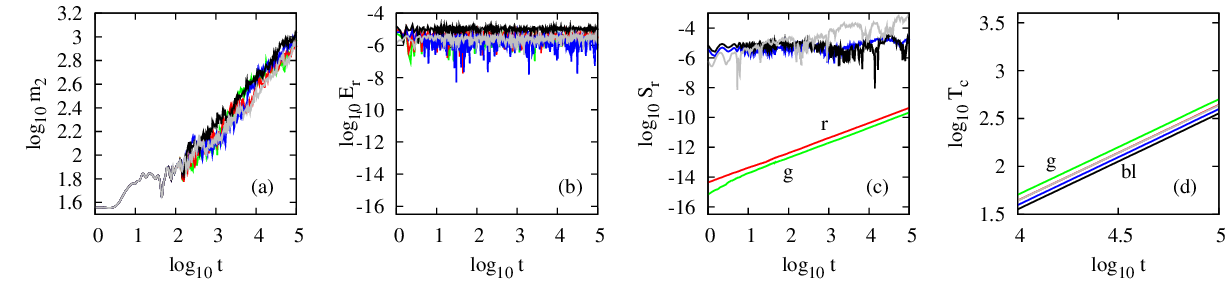}
\caption{(Color online) Results for the integration of
    $H_D$ (\ref{HDDNLS}) by the fourth order SI ABC$^4_{[\text{Y}]}$
    for $\tau=0.05$ [(gy) grey], and the fourth order SIs
    SIFT$^4_{864}$ for $\tau=0.25$, SIFT$^4_{1064}$ for $\tau=0.25$,
    SS$^4_{864}$ for $\tau=0.25$ and SS$^4_{1064}$ for $\tau=0.25$
    [(r) red; (g) green; (bl) black; (b) blue].  The panels are as in
    Fig.~\ref{f1}. Note that in panel (d) the red and grey curves
    practically overlap.}
\label{f3}
\end{figure*}

Therefore,  we compare in Fig.~\ref{f3} this integrator with the
  remaining fourth order schemes that we consider in our study:
  SIFT$^4_{864}$, SIFT$^4_{1064}$, SS$^4_{864}$ and
  SS$^4_{1064}$. Again all schemes accurately reproduce the evolution
  of $m_2$ (Fig.~\ref{f3}a) and keep the relative energy error
  practically constant, i.e.~$E_r\approx 10^{-5}$
  (Fig.~\ref{f3}b). The SIFT$^4_{864}$ and SIFT$^4_{1064}$ methods,
  which implement Fourier transforms, have again small $S_r$ values,
  while SS$^4_{864}$ and SS$^4_{1064}$ methods preserve the norm quite
  accurately as they produce (larger) $S_r$ values, which nevertheless
  remain practically constant (Fig.~\ref{f3}c). The good behavior of
  the SS$^4_{864}$ and SS$^4_{1064}$ integrators is probably due to
  the fact that the corresponding fourth order ABAH864 and ABAH1064
  integrators, on which they are based, also eliminate some higher
  order terms.

From Fig.~\ref{f3}d we see that all methods considered in
  Fig.~\ref{f3} require more or less similar CPU times, with the
  SS$^4_{864}$ scheme showing the best performance. Nevertheless, one
  should be more careful about the significance of CPU time
  improvements.  From the results of Fig.~\ref{f3} we see that using
  the SS$^4_{864}$ with $\tau=0.25$ we need $\sim1.2$ times less CPU
  time than the ABC$^4_{[\text{Y}]}$ with $\tau=0.05$, which is the
  best performing scheme among the ones considered in Figs.~\ref{f1}
  and \ref{f2}. Comparing the SS$^4_{864}$ method with the SS$^2$ and
  SIFT$^2$ methods usually used in numerical studies of the DDNLS
  model \cite{FKS09,SKKF09,LBKSF10,BLSKF11,BLGKSF11} we see that the
  gain factor increases even more. In particular, SS$^4_{864}$ scheme
  requires $\sim1.4$ and $\sim2.0$ times less CPU time than the
  SIFT$^2$ with $\tau=0.05$ and the SS$^2$ with $\tau=0.02$
  respectively (Fig.~\ref{f1}).
Although one might argue that these CPU time gain factors are
not too big, we should keep in mind that long time simulations up to
$t=10^7-10^8$ of the DDNLS model with $N\sim1000$ sites could require
(depending on the particular computer used) up to $\sim10$ days of
computations. Thus a gain factor of 2 is practically significant
as it can considerably reduce the computation time.

\begin{figure*}[t]
\includegraphics[width=\textwidth, keepaspectratio]{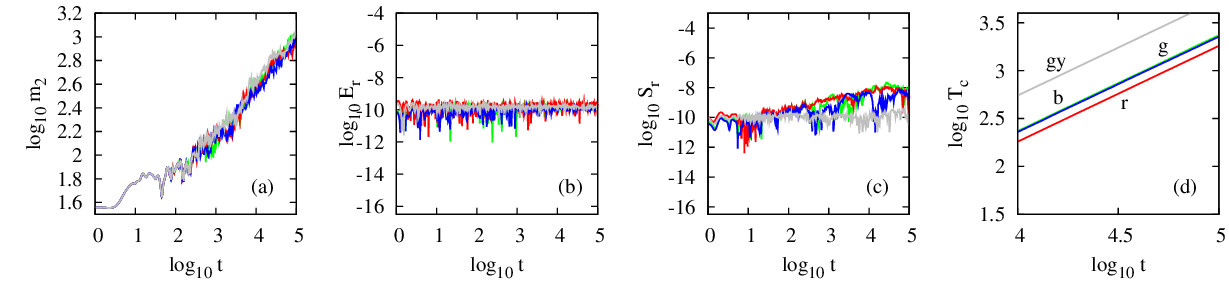}
\caption{(Color online) Results for the integration of
    $H_D$ (\ref{HDDNLS}) by the fourth order SI SS$^4_{864}$ for
    $\tau=0.015625$ [(gy) grey], and the sixth order SIs
    ABC$^6_{[\text{Y}]}$ for $\tau=0.03$, ABC$^6_{[\text{SS}]}$ for
    $\tau=0.125$ and ABC$^6_{[\text{KL}]}$ for $\tau=0.04$ [(g) green;
      (r) red; (b) blue].  The panels are as in Fig.~\ref{f1}. Note
    that in panel (d) the green and blue curves practically overlap.}
\label{f4}
\end{figure*}

To keep $E_r \approx 10^{-5}$ 
most of the studied SIs of order higher than four require
  large integration steps, which are already outside the
stability domain of these algorithms. In order to avoid this situation we lowered
  the relative energy error to $E_r \approx 10^{-10}$ for the
  comparative study of these methods. From the results of
  Fig.~\ref{f4} we see that, as expected, the sixth order SIs
  ABC$^6_{[\text{Y}]}$, ABC$^6_{[\text{SS}]}$ and
  ABC$^6_{[\text{KL}]}$ are more efficient than the SS$^4_{864}$ which
  showed the best performance among all integration schemes of
  Figs.~\ref{f1}--\ref{f3}, as they correctly reproduce the evolution
  of $m_2$ (Fig.~\ref{f4}a), keep bounded both the energy
  (Fig.~\ref{f4}b) and the norm (Fig.~\ref{f4}c) relative errors
  (although a slight increase is observed for $S_r$) and require less
  CPU time (Fig.~\ref{f4}d).  

\begin{figure*}[t]
\includegraphics[width=\textwidth, keepaspectratio]{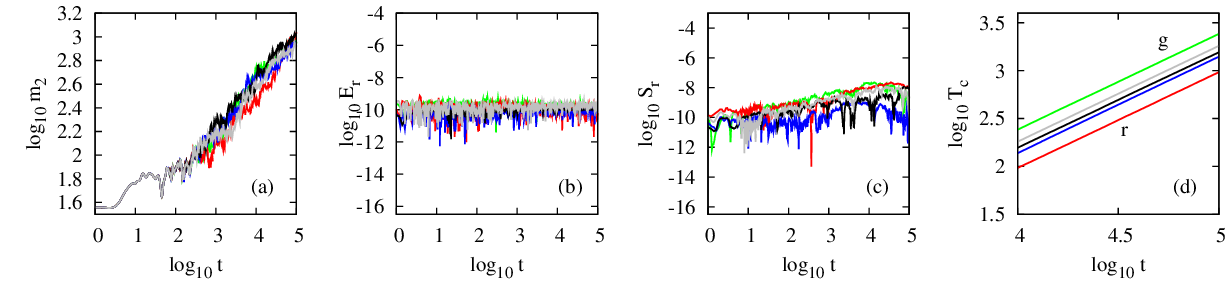}
\caption{(Color online) Results for the integration of
    $H_D$ (\ref{HDDNLS}) by the sixth order SI ABC$^6_{[\text{SS}]}$
    for $\tau=0.125$ [grey], the eight order SIs ABC$^8_{[\text{Y}]}$ for
    $\tau=0.0625$, ABC$^8_{[\text{SS}]}$ for $\tau=0.2$,
    ABC$^8_{[\text{KL}]}$ for $\tau=0.125$ [(g) green; (r) red; (b)
      blue], and the tenth order SI ABC$^{10}_{[\text{SS}]}$ for
    $\tau=0.2$ [(bl) black].}
\label{f5}
\end{figure*}

 Implementing SIs of even higher order we obtain methods
  with even better performances (namely schemes ABC$^8_{[\text{Y}]}$,
  ABC$^8_{[\text{SS}]}$, ABC$^8_{[\text{KL}]}$ and
  ABC$^{10}_{[\text{SS}]}$) than ABC$^6_{[\text{SS}]}$
  (Fig.~\ref{f5}). Nevertheless, only the increase of the SI's order
  is not sufficient to guarantee improvement of the computational
  behavior, as the simultaneous growth of steps could augment the CPU
  time requirements. For instance, this is why ABC$^8_{[\text{Y}]}$
  and ABC$^{10}_{[\text{SS}]}$ require more CPU time than
  ABC$^6_{[\text{SS}]}$ and ABC$^8_{[\text{SS}]}$ respectively
  (Fig.~\ref{f5}d).  

Our results indicate that the construction of efficient triple split
SIs can allow the integration of the DDNLS for longer times, and
numerically tackle questions about the asymptotic behavior of wave
packets.  We note that the ABC$^8_{[\text{SS}]}$ SI
  required the less CPU time among all tested schemes
  (Fig.~\ref{f5}d).

\section{Conclusions and Discussion}\label{sec:con}

In summary, we presented ways to use SIs for Hamiltonian systems that
do not split in two integrable parts, as traditional symplectic
methods require, but in three. For such systems we 
  considered several high order three part split SIs
  based on already developed composition methods and emphasized their
practical importance.  In particular, we showed that such three part
split SIs are more efficient numerical schemes than other symplectic
and non-symplectic methods in terms of both accuracy and CPU time
requirements. These characteristics are of particular importance for
the long time integration of multidimensional systems like the DDNLS
model, whose asymptotic behavior is currently a very debatable issue.

 Many of the studied integration schemes showed a quite
   satisfactory behavior with respect to both their accuracy and
   efficiency. For example integrator SS$^4_{864}$ required the least
   CPU time among all tested schemes of order up to four and kept
   practically constant also the relative error of the system's second
   integral of motion i.e.~its norm. In addition, all algorithms based
   on the integration of the $\mathcal{B}=B+C$ part of Hamiltonian
   (\ref{HDDNLS}) via Fourier transforms, i.e.~methods SIFT$^2$,
   SIFT$^4$, SIFT$^4_{864}$ and SIFT$^4_{1064}$ succeeded in keeping
   the relative error $S_r$ very low (although it increased with
   integration time). A drawback of these methods is that, due to the
   applications of Fourier transforms, they require the number of
   lattice sites to be $2^k$, $k \in \mathbb{N}^*$, although this is
   not always the case in numerical simulations. Also schemes referred
   as ABC methods, which are based on the fact that the studied
   Hamiltonian (\ref{HDDNLS}) is split in exactly three integrable
   parts, proved to be quite efficient methods, whose performance
   generally improve with increasing order.

We hope that our results will draw the interest of the community in
the construction of three part split SIs, and will initiate future
research both for the theoretical development of new, improved
integrators of this type, as well as for their applications to
different dynamical systems. Keeping in mind that such SIs can provide
efficient numerical schemes for the long time integration of
Hamiltonian systems with many degrees of freedom (like the DDNLS
model), it would be interesting to investigate if the possible
addition of a corrector term can improve their accuracy, as done for
traditional two part split methods (see e.g.~\cite{LR01}).

\section*{Acknowledgments}
We thank the anonymous referee for many valuable
  suggestions that helped us to greatly improve our paper.
Ch.S.~would like to thank S.~Anastasiou, G.~Benettin and J.~Laskar for
useful discussions, as well as the Max Planck Institute for the
Physics of Complex Systems in Dresden for its hospitality during his
visits in 2012 and 2013, when part of this work was carried
out. Ch.S.~was supported by the Research Committees of the Aristotle
University of Thessaloniki (Prog. No 89317) and the
  University of Cape Town (Start-Up Grant, Fund No 459221), as well
  as by the European Union (European Social Fund - ESF) and Greek
national funds through the Operational Program ``Education and
Lifelong Learning'' of the National Strategic Reference Framework
(NSRF) - Research Funding Program: ``THALES. Investing in knowledge
society through the European Social Fund''. S.E. acknowledges the
support of the European Union Seventh Framework Program
(FP7/2007-2013) under grant agreement no. 282703. G.E.~would like to
thank P.~Jung for fruitful discussions.





\end{document}